\documentclass[twocolumn]{aastex631}

\usepackage{bm}

\newcommand{\msun}{{M}_{\sun}}
\usepackage{bbding}
\usepackage{lineno}

\shorttitle{}
\shortauthors{Xia et al.}

\begin{document}

\title{Strong Amplitude Modulation of Hard-band X-ray QPO with Soft-band Flux in RE J1034+396}

\author[0009-0005-3916-1455]{Ruisong Xia}
\author[0000-0001-5525-0400]{Hao Liu\textsuperscript{\Envelope}}
\author[0000-0002-1935-8104]{Yongquan Xue\textsuperscript{\Envelope}}

\affiliation{CAS Key Laboratory for Research in Galaxies and Cosmology, Department of Astronomy, University of Science and Technology of China, Hefei 230026, China;  liuhao1993@ustc.edu.cn, xuey@ustc.edu.cn}
\affiliation{School of Astronomy and Space Science, University of Science and Technology of China, Hefei 230026, China}

\begin{abstract}
The X-ray quasi-periodic oscillation (QPO) is a remarkable form of variability in systems of compact object accretion.
RE J1034+396, harboring the most significant X-ray QPO {in active galactic nuclei (AGNs)}, is the most noteworthy source for in-depth analysis of {AGN} X-ray QPO properties.
A long-term evolution of {its} QPO has been observed over the course of the observations.
However, the short-term variability of {its} QPO properties remains {unexplored} within each observation that has long good time intervals (GTIs).
We collect 12 XMM-Newton observations of RE J1034+396 with GTIs longer than 60 ks from publicly available data and conduct a detailed wavelet analysis focusing on the short-time modulation of the QPO.
The QPO signals are found to undergo amplitude modulation in both the soft and hard bands, with a typical timescale of 17 ks.
The soft flux is significantly higher when the hard QPO is present. They are highly correlated, with an average cross-correlation function (CCF) peak coefficient of 0.61 and a lag of approximately 3 ks.
This novel finding provides fresh insights into the potential connection between the components of the corona emitting soft and hard X-ray photons.
The CCF lag between the soft flux and the hard QPO evolves across the observations, potentially sharing the same origin as the previously observed interconnected evolution between QPO frequency and time lag.

\end{abstract}


\section{INTRODUCTION}\label{sec1}
The X-ray quasi-periodic oscillation (QPO) is a remarkable variability in systems with compact object accretion. Over the last two decades, X-ray QPOs have been observed in several active galactic nuclei (AGNs), providing an important path for studying accretion physics \citep[e.g.][]{2008Natur.455..369G,2008ApJ...679..182E,2016ApJ...819L..19P,2017ApJ...849....9Z,2015MNRAS.449..467A,2023MNRAS.523L..26K}. 
RE J1034+396, harboring the most significant X-ray QPO {in AGNs}, is the most noteworthy source for in-depth analysis of {AGN} X-ray QPO properties.

QPO signals have been detected in all XMM-Newton observations for RE J1034+396 with good time intervals (GTIs) longer than 60 ks \citep[e.g.][]{2008Natur.455..369G, 2020MNRAS.495.3538J, 2024ApJ...961L..32X}. However, QPO signals are less detected in those with shorter GTIs \citep{2020MNRAS.495.3538J}. It is possible that the QPO signals are intermittent on short timescales rather than constantly present throughout an observation longer than 60 ks.

The power spectral density (PSD) is widely used to analyze the properties of the QPO signals. However, PSD cannot capture the short-term variability of QPO signals due to its inherent assumption of stationarity.  Wavelet analysis provides better localization in both time and frequency domains, allowing for more detailed analysis of non-stationary signals where frequency and amplitude vary over time \citep{1998BAMS...79...61T, 2009arXiv0906.4176L}.
Moreover, the wavelet coherence transform provides abundant information about the coherence and time lag between the soft and hard bands of the X-ray variability \citep{2023MNRAS.526.3441W}. Similar to how the PSD describes the power distribution of a single time series across frequencies, and {how} the cross-spectral density illustrates the interaction between two time series, the wavelet coherence transform extends these concepts to a time-frequency framework.

Wavelet transform has been used to search for potential X-ray QPOs for AGNs, e.g., 3C~273 \citep{2008ApJ...679..182E}, 
PKS~2155$-$304 \citep{2009A&A...506L..17L}, MCG$-$06$-$30$-$15 \citep{2018A&A...616L...6G}, 1H~0707$-$495 \citep{2018ApJ...853..193Z}, ESO~113$-$G010 \citep{2020ChA&A..44...32Z}, NGC~1365 \citep{2024arXiv240516187Y}, {and RE J1034+396 }\citep{2023MNRAS.524.1478G}. It is quite efficient at capturing variable signals and analyzing their amplitude modulation (AM) and frequency modulation (FM).
AM and FM are two key types of modulation used in the study of QPOs. AM refers to the process where the amplitude of the QPO signal varies over time, while FM, on the other hand, involves changes in the frequency of the QPO signal over time.
The FM of the QPO signals in RE J1034+396 was found by \citet{2010A&A...524A..26C} and thought to be correlated with the flux.
 
This work aims to further study the short-term variability of the QPO properties in RE J1034+396 to provide insights into the corona structure and variability.
We collect 12 XMM-Newton observations of RE J1034+396 from publicly available data (Section~\ref{sec2}) and conduct a detailed wavelet analysis (Section~\ref{sec3}). We discuss the types, origin, and long-term evolution of the modulation in Section~\ref{sec4}, and the conclusions are presented in Section~\ref{sec5}.

\section{Data and Reduction}\label{sec2}
\begin{table*}
\begin{center}
\caption{Information of \textit{XMM-Newton} Observations. }\label{tab1}

\begin{tabular}{ccccccc}
\hline
\hline
Sequence &Obs. No. & ObsID & Obs. Date & GTI & $F_{\rm QPO}$ & $F_{+}$\\
&&&(yyyy-mm-dd)&($\rm ks$)&(\%)&(\%)\\
(1)&(2)&(3)&(4)&(5)&(6)&(7)\\
\hline
 1 & Obs-b & 0824030101 & 2018-10-30 & $64.7 $ & 64.57/90.24 & 91.91 \\
 2 & Obs-5 & 0865011301 & 2021-04-24 & $91.4 $ & 62.80/63.56 & 77.44 \\
 3 & Obs-6 & 0865011401 & 2021-05-02 & $87.2 $ & 56.31/83.72 & 93.92 \\
 4 & Obs-7 & 0865011501 & 2021-05-08 & $91.4 $ & 26.75/63.38 & 87.83 \\
 5 & Obs-8 & 0865011601 & 2021-05-12 & $88.8 $ & 39.14/81.26 & 88.69 \\
 6 & Obs-9 & 0865011701 & 2021-05-16 & $91.5 $ & 10.95/87.85 & 71.91 \\
 7 & Obs-1 & 0865010101 & 2020-11-20 & $86.4 $ & 0.00/72.34 & 60.09 \\
 8 & Obs-10 & 0865011801 & 2021-05-30 & $85.5$ & 13.77/76.79 & 50.65 \\
 9 & Obs-2 & 0865011001 & 2020-11-30 & $85.2 $ & 40.96/81.22 & 28.76 \\
 10 & Obs-3 & 0865011201 & 2020-12-02 & $89.2 $ & 60.48/88.36 & 11.64 \\
 11 & Obs-4 & 0865011101 & 2020-12-04 & $91.0 $ & 71.97/84.75 & 16.48 \\
 12 & Obs-a & 0506440101 & 2007-05-31 & $79.5 $ & 79.87/67.94 & 44.64 \\
\hline
\hline

\end{tabular}

\tablecomments{Columns are as follows: 
(1) The hypothetical sequence of the interconnected evolution for the observations \citep{2024ApJ...961L..32X};
(2) The observation number in this paper; 
(3) The observation ID; 
(4) The observation date; 
(5) The good time interval in EPIC-pn; 
(6) The fraction of time when the soft/hard QPO is present; 
(7) The fraction of time when the hard QPO lags behind the soft QPO ({i.e., }the QPO time lag is positive).}
\end{center}
\end{table*}

We collect all light curves presenting QPO signals {with GTIs} longer than 60 ks, observed by the \textit{XMM-Newton} satellite \citep{2001A&A...365L...1J} since May 31, 2007{, when the QPO was first observed}. Twelve observations are selected for this study and listed in Table~\ref{tab1}, arranged according to their phases in a hypothetical interconnected evolutionary cycle \citep[see Figure~4 in][]{2024ApJ...961L..32X}, beginning at Obs-b and ending at Obs-a. 
We process the European Photon Imaging Cameras (EPIC) data using the Science Analysis Software (SAS, version 20.0.0) with the latest calibration files. To avoid contamination from {periods of flaring in the particle background}, we exclude data with rate values {higher} than 0.4 counts per second in the 10–12 keV band.

We extract the EPIC-pn light curves for all 12 observations in the soft (0.3--1 keV) and hard (1--4 keV) bands. Using the \textsc{evselect} and the \textsc{epiclccorr} {tasks}, we obtain the final background-subtracted light curves in the two bands. The binning time for the light curves is set as 100 seconds. 
{The light curves are linearly interpolated during background flare periods that account for approximately 2.2\% of the total exposure time.} Source regions are selected within a $40 \arcsec$ radius circle centered on the source for all observations, except for the observation Obs-a. Due to a significant pile-up effect in Obs-a, we choose an annulus source region with inner and outer radii of $10 \arcsec$ and $40 \arcsec$, respectively.

\section{Analysis and Results}\label{sec3}
\subsection{Wavelet and Wavelet Coherence Transform}
\begin{figure*}
    \centering
    \includegraphics[scale=0.75, angle=-90]{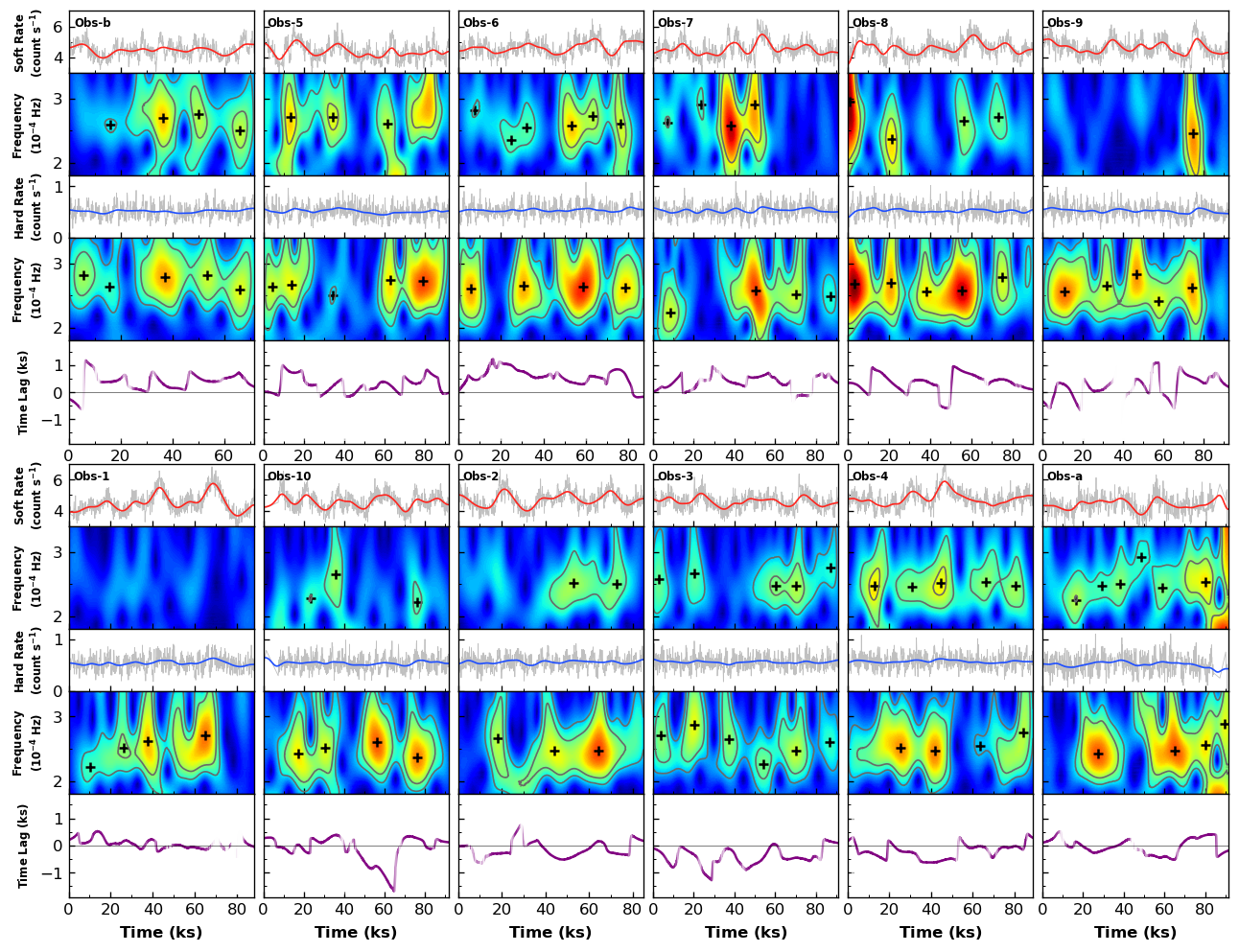}
    \caption{Light curves, wavelet maps, and time lag as a function of time for the 12 observations in the soft and hard bands. The light curves are plotted in gray. The red (for the soft band) and blue (for the hard band) curves are smoothed light curves to showcase variability excluding the influence of the QPO itself. Wavelet maps are highlighted with gray contours representing 90\% and 99\% confidence levels. The localized peaks for individual QPO patterns are marked by black cross symbols in the wavelet maps. The time lag is estimated using wavelet coherence transform, with {positive values representing hard-band lags, and the opacity of the purple curve represents coherence}. Observations are arranged according to their phases within the hypothetical long-term cycle.}
    \label{fig1}
\end{figure*}

We use the open-source wavelet transform software \textsc{PyWavelets} \citep{lee_pywavelets_2019}, adopting the Complex Morlet wavelet with $2\pi\nu_0=6$ {(where $\nu_0$ is the center frequency)} as a canonical choice \citep{2005MNRAS.361..645L}, to analyze the soft- and hard-band light curves, as shown in Figure~\ref{fig1}. The frequency range for displaying the wavelet map is set between $1.8\times 10^{-4}\ \rm Hz$ and $3.4\times 10^{-4}\ \rm Hz$.
Similar to \citet{2010A&A...524A..26C}, we determine the confidence levels via Monte Carlo simulations. We fit a simple power law to the power spectral density (PSD) of the light curves and generate 2000 mock light curves based on this power law using \textsc{astroML} \citep{astroML}. 
{While a broken or bending power law might represent a common modeling choice \citep{2014MNRAS.445L..16A}, \citet{2020MNRAS.495.3538J} have demonstrated that the power law and bending power law models show no statistically significant difference. Based on this evidence and considering both simplicity and effectiveness, we maintain that the power law represents a reasonable choice for our analysis.} 
By comparing the wavelet maps of these mock light curves with the real light curves, we determine the confidence levels, marking the 90\% and 99\% levels as contours. 
We determine when QPO is present or absent in both the soft and hard bands by assessing whether the confidence level exceeds 90\% in the wavelet map within the QPO frequency range.
The fraction of time when the soft or hard QPO is present for each observation is detailed in column (6) of Table~\ref{tab1}, and also seen in the lower two panels (e) and (f) of Figure~\ref{fig4} for their evolution in the hypothetical sequence of the observations.
The overall trend in the evolution of the QPO frequency is evident in the wavelet maps.

Combining the source codes of \textsc{PyCWT}\footnote{Sebastian Krieger and Nabil Freij. PyCWT: wavelet spectral analysis in Python. V. 0.4.0-beta. Python. 2023. https://github.com/regeirk/pycwt.} and \textsc{PyWavelets}, we conduct wavelet coherence transform for the light curves of the soft and hard bands of each observation keeping the same parameters of the Complex Morlet {stated above}. This process yields both a wavelet coherence phase angle map and a wavelet coherence amplitude map. We sum the wavelet coherence phase angle map between $2\times 10^{-4}\ \rm Hz$ and $3\times 10^{-4}\ \rm Hz$ and calculate the QPO time lag by $\tau=\phi/2\pi f$ \citep{2014A&ARv..22...72U, 2023MNRAS.526.3441W}, where $\tau$ is the time lag, $\phi$ is the phase angle, and $f$ is the corresponding frequency in the map. The coherence is similarly estimated with the wavelet coherence amplitude map. The time lag as a function of time is shown as purple curves in Figure~\ref{fig1}, {where a higher opacity indicates a greater coherence}. The fractions of time when the QPO time lag is positive for each observation are also detailed in Table~\ref{tab1}, and also seen in panel (b) in Figure~\ref{fig4} for its long-term evolution in the hypothetical sequence of the observations.

\subsection{Intermittent QPO}\label{sec32}
\begin{figure}
    \centering
    \includegraphics[scale=0.46]{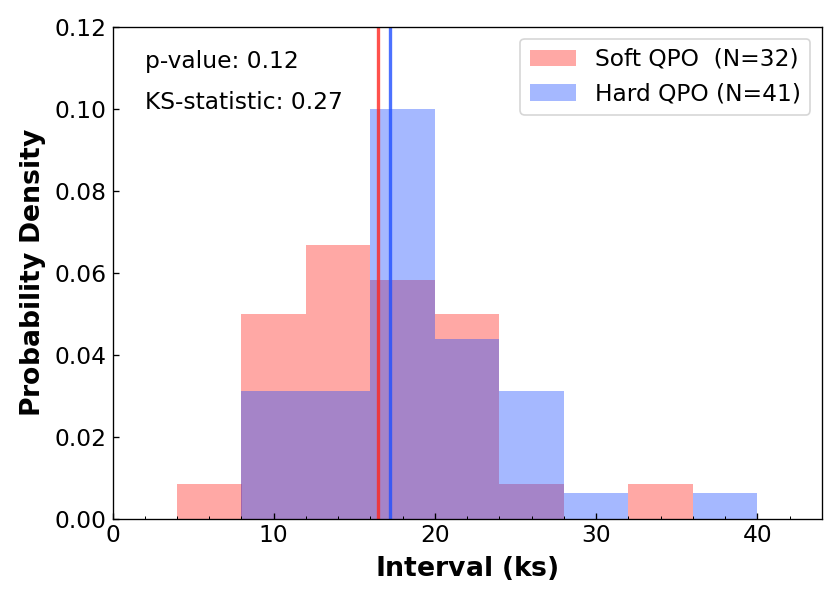}
    \caption{The distribution of the intervals of the peaks of the QPO patterns in the wavelet map for the soft (red) and the hard (blue) QPOs. The solid lines indicate the medians of the distributions. The p-value, indicating the probability of observing the data assuming the null hypothesis that the two distributions are the same, and the KS-statistic, indicating the magnitude of difference between the two distributions, are presented in the upper left corner. The total counts of the intervals $N$ are labeled in the upper right corner.}
    \label{fig2}
\end{figure}

\begin{figure*}
    \centering
    \includegraphics[scale=0.8]{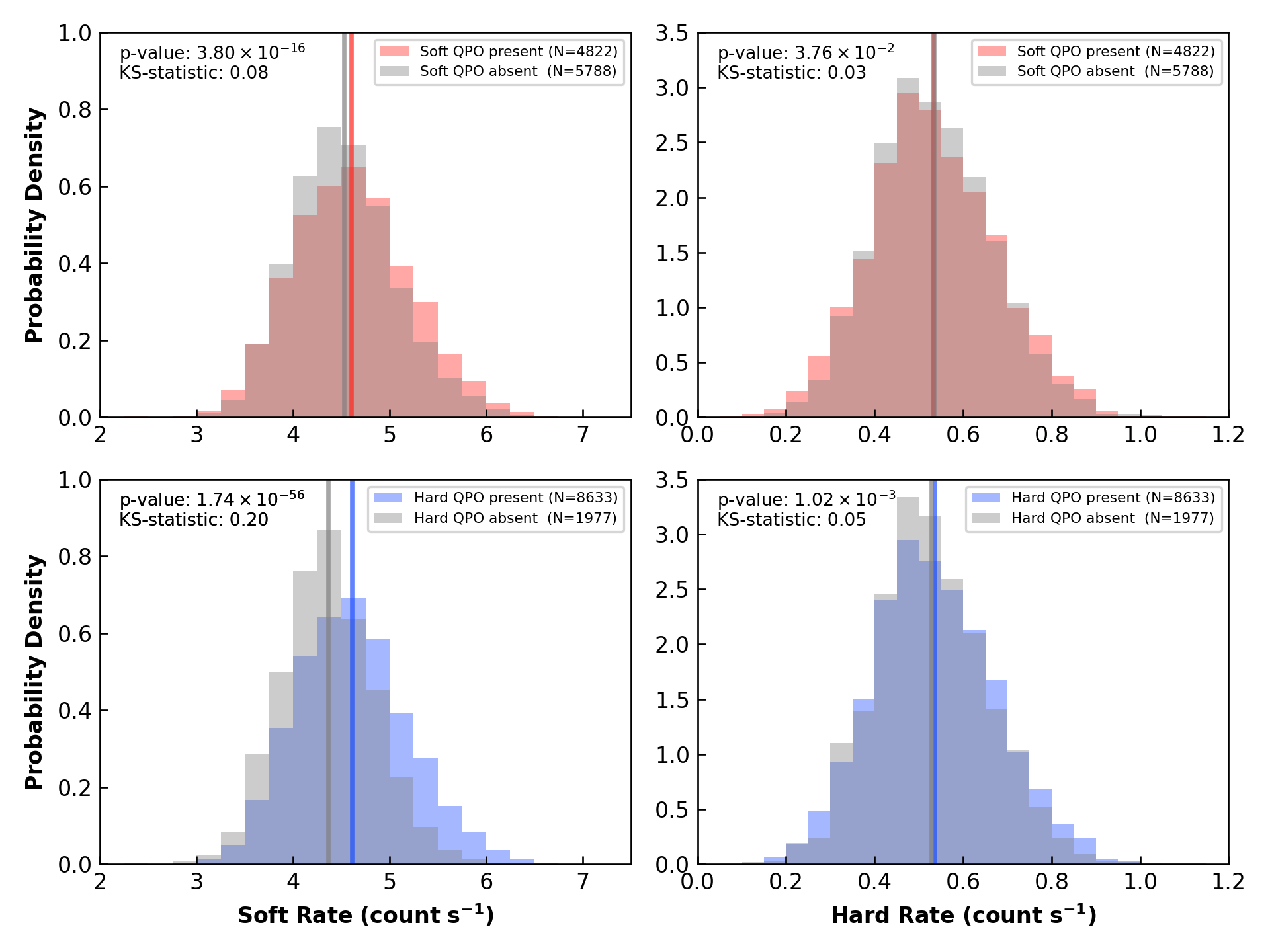}
    \caption{The distributions for the count rates of the soft (left panels) and hard bands (right panels) when the soft (red; upper panels) or hard (blue; lower panels) QPO is present (colored) or absent (grey). The p-value, indicating the probability of observing the data assuming the null hypothesis that the two distributions are the same, and the KS-statistic, indicating the magnitude of difference between the two distributions, are presented in the upper left corner of each panel. The total counts of the light curves $N$ are labeled in the upper right corner of each panel.}
    \label{fig3}
\end{figure*}

The QPO patterns are intermittent in the wavelet maps, as shown in Figure~\ref{fig1}. We use \textsc{maximum\_filter} from \textsc{scipy} to determine the localized peaks of the wavelet map, which represent individual patterns of the QPOs. The localized peaks that are outside the QPO range (less than 90\% confidence level) have been removed. We estimate the intervals of these localized peaks and build histograms (Figure~\ref{fig2}). 
A Kolmogorov-Smirnov test (KS-test) is used to examine the distributions of the intervals of the QPO pattern peaks in these two {bands}. The p-value of 0.12 suggests a considerable likelihood that the two distributions are the same. 
If we use T-test instead, only considering the likelihood of the mean values being the same, the p-value will be 0.48, which is much higher.
To avoid potential omissions when detecting localized peaks, we use the median value rather than the mean to represent the typical interval time. This median interval and the standard deviation is $16.9\pm 7.3$ ks ($16.5 \pm 8.0$ ks for the soft QPO and $17.2 \pm 6.7$ ks for the hard QPO). This indicates that the QPOs are likely undergoing AM with a {timescale} of approximately 17 ks, corresponding exactly to the $\rm (5.25\pm 1.86) \times 10^{-5}\ Hz$ oscillation of the soft light curve (see details in {Appendix}~\ref{secA} and Figure~\ref{figA1}).

\subsection{Connection between QPO and Count Rates}

We construct histograms of the soft and hard count rates for cases where the QPO is present {or} absent and perform a KS-test for each case (Figure~\ref{fig3}). Note that since the total counts of the light curves are quite large (e.g., a 90 ks light curve provides 900 data points for statistical analysis, and we have a total of 12 observations), the tiny deviation between the distributions will be prominent, resulting in small p-values. We observe that, except for the hard count rate distributions when the soft QPO is considered (upper right panel in Figure~\ref{fig3}), the count rate distributions differ significantly {(p-value less than 0.01)} depending on the presence of the QPO. This difference is especially notable for the soft count rate distribution when comparing the presence and absence of the hard QPO (lower left panel in Figure~\ref{fig3}), indicating that the soft flux level is higher when the hard QPO is present. 
A similar analysis of the QPO time lag and count rates is presented in Appendix~\ref{secC} and Figure~\ref{figC1}, though the correlation is weaker.

\begin{figure*}
    \centering
    \includegraphics[scale=0.8]{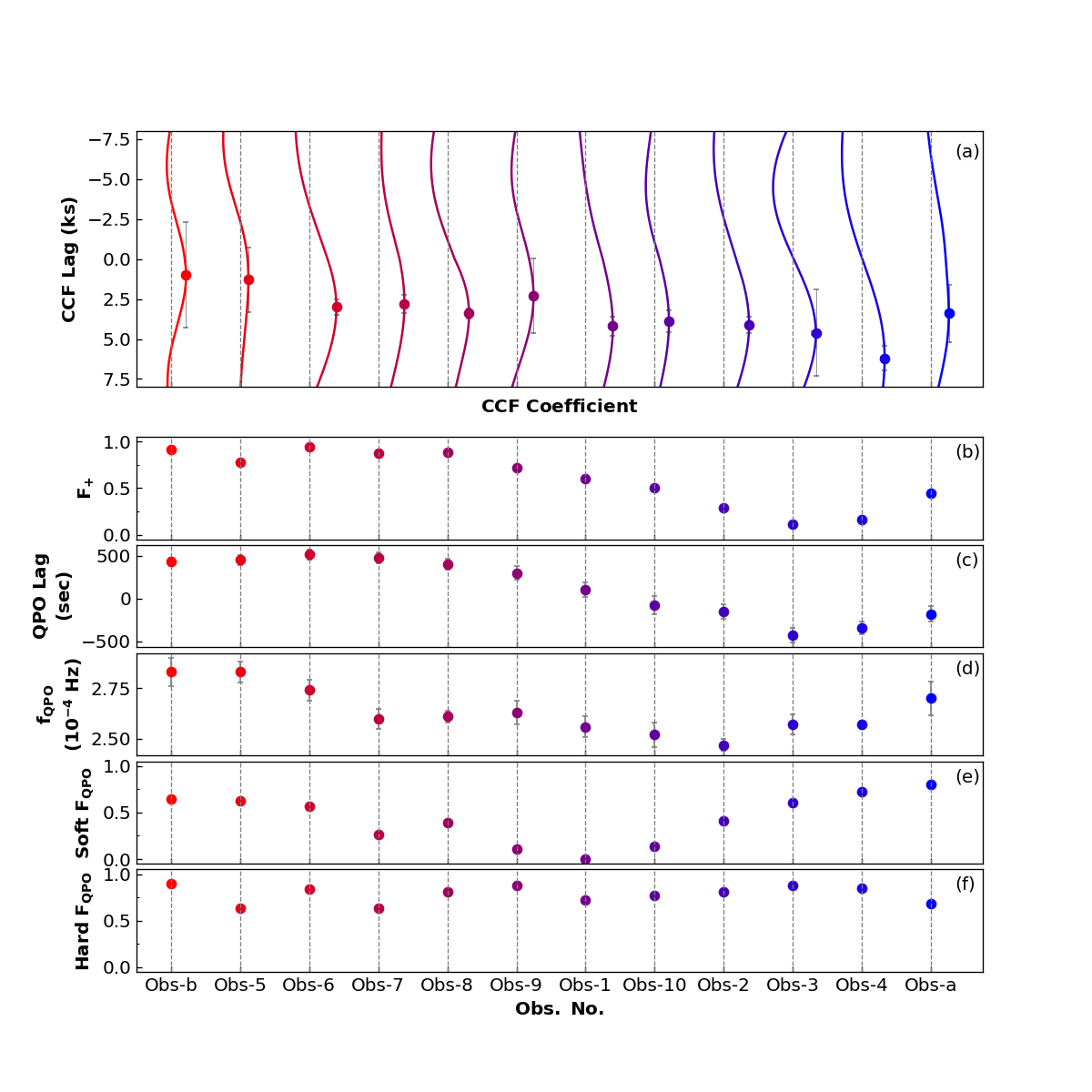}
    \caption{Panel (a) shows the normalized CCFs of the soft flux (represented by the smoothed count rate in the soft band) and the hard QPO amplitude (represented by the integrated wavelet map in the QPO frequency range). The vertical dashed lines denote 0 values of the CCF coefficient for each observation, with positive values relative to these dashed lines indicating that the soft flux lags behind the hard QPO amplitude. The spacing between the dashed lines corresponds to a coefficient of 2. The colors transition from red to blue along the hypothetical sequence of the observations and the positive CCF peaks are marked by colored circles. 
    Panels (c) and (d) show the QPO frequency ($f_{\rm QPO}$) and QPO time lag estimated by \citet{2024ApJ...961L..32X}.
    Panels (b), (e), and (f) display the evolution of the fraction of time when the QPO time lag is positive ($F_+$), the presence of the soft QPO (soft $F_{\rm QPO}$), and the presence of the hard QPO (hard $F_{\rm QPO}$).
    The observations are arranged under the hypothetical sequence.}
    \label{fig4}
\end{figure*}

The intensity of the soft band is correlated with the presence of the hard-band QPO, prompting further investigation of this relationship. We sum the amplitude of the wavelet map between $2\times 10^{-4}\ \rm Hz$ and $3\times 10^{-4}\ \rm Hz$ for each time bin of the light curves as an indicator of QPO amplitude. Subsequently, we calculate the normalized cross-correlation function (CCF) of the soft flux (represented by the smoothed count rate in the soft band) and the hard QPO amplitude (represented by the integrated wavelet map in the QPO frequency range). 
{The uncertainties are estimated using the bootstrap method with 1000 resamplings based on the errors of the light curves, which are calculated as the standard deviation of the parameter distribution.}
The result is plotted in panel (a) of Figure~\ref{fig4}. Note that the count rates used for calculating CCF have been smoothed to mitigate the influence of the QPO itself, as plotted in the red (soft) and blue (hard) curves in Figure~\ref{fig1}. 
We mark the positive CCF peaks (indicating that the soft flux lags behind the hard QPO amplitude) by circles and an obvious evolution trend under the hypothetical sequence for the observations is shown (Figure~\ref{fig4}), as the CCF lag is almost consistently decreasing, except for Obs-a, indicating the same origin as the interconnected evolution between QPO frequency and QPO time lag \citep{2024ApJ...961L..32X}.

\section{Discussions}\label{sec4}
\subsection{AM and FM}

{While \citet{2023MNRAS.524.1478G} have suggested that QPO fluctuations in wavelet maps could be intrinsic to the stochastic nature of Lorentzian signals, evidence for} a temporal modulation of the QPO on a 24 ks timescale has been detected by \citet{2010A&A...524A..26C} using the analysis of variance method \citep{1996ApJ...460L.107S} on the QPO period curve of obs-a, which could be an FM. This is close to the median intervals of {$16.9\pm 7.3$} ks in this work (Section~\ref{sec32}).

This temporal modulation of the QPO period (FM) was found to be correlated with the flux.
However, considering 12 observations with {a} cumulative exposure longer than 1000 ks, the CCFs revealed a maximum correlation of $0.61\pm0.13$ between the hard QPO amplitude and the soft flux, which is more significant than the maximum correlation of $0.37\pm0.14$ between the hard QPO frequency and the soft flux (see details in {Appendix}~\ref{secB} and {Figures}~\ref{figB1}~and~\ref{figB2}). This indicates that although both AM and FM exist in the QPO, the {soft} flux is more correlated with the AM, and thus with the presence of the {hard} QPO.

\subsection{Origin of Connection between Hard QPO and Soft Flux }

By carefully examining the smoothed light curves and the wavelet map in Figure~\ref{fig1}, we have inferred a trend: when the QPO pattern appears in the hard band, the soft flux level shows a strong increase with a lag, while the hard flux remains quite stable. There should be a connection between the possible components of the corona corresponding to the soft and hard bands and the influence of the QPO on the hard flux is somehow mitigated.

RE J1034+396 was recently reported to have ultra-fast outflows (UFOs), which appears to be correlated with the flux, with a higher flux state corresponding to a more significant UFO \citep{2024arXiv240507494X}. This suggests a potential correlation between the QPO and the UFO.
The spiral wave model, based on the accretion-ejection instability (AEI), is the one that correlates with outflows \citep{2002A&A...387..497V, 2019NewAR..8501524I}. In this model, the QPO frequency is typically $0.1\text{--}0.3 f_{int}$, where $f_{int}$ is the rotation frequency at the inner radius of the disk \citep{2002A&A...387..497V}. Considering the central black hole mass of $\sim 10^6 \text{--} 10^7\msun$  \citep{2016A&A...594A.102C} for RE J1034+396, the AEI-modeled QPO frequency falls within the range of {$(0.22\text{--}6.60) \times 10^{-4} \ \rm{Hz}$}, which is compatible with the detected QPO frequency. Additionally, this model would explain the unstable but long-lived QPO and it launches Alfv\'en waves carrying energy and momentum from the disk to the corona, possibly being the source of an outflow \citep{2002A&A...387..497V, 2019NewAR..8501524I}. 
This model could be a promising explanation for the origin of the QPO, as all these features match the properties observed in RE J1034+396, even though it was originally proposed to explain black hole binaries.

Assuming two warm Comptonization components of the corona \citep{2021MNRAS.500.2475J} emitting the soft and hard X-rays, respectively, while the QPO originates from the disk, and the photons are up-scattered by the corona and the QPO time lag is due to the light propagation timescale of about hundreds of seconds \citep{2020MNRAS.495.3538J, 2024ApJ...961L..32X}. The Alfv\'en waves should affect only the soft component for RE J1034+396, transferring energy from the disk to the soft component of the corona and increasing the soft flux with a lag of approximately 3~ks behind the hard QPO. In this scenario, Alfv\'en waves could become the connection between the hard QPO and the soft flux.
{An in-depth analysis incorporating both the outflow characteristics and the QPO evolution will help validate these conjectures.}

\subsection{Long-term Evolution}
In Figure~\ref{fig4}, we present the evolution of the CCF lag, the QPO frequency ($f_{\rm QPO}$) and time lag \citep{2024ApJ...961L..32X}, the fraction of time when the QPO time lag is positive ($F_+$), and the presence of the soft and hard QPO (soft or hard $F_{\rm QPO}$), along the hypothetical sequence of the observations. We observe that, except for the hard $F_{\rm QPO}$, which appears to evolve randomly, the other properties demonstrate a similar trend in relation to the QPO frequency or the QPO time lag as reported by \citet{2024ApJ...961L..32X}, possibly also with hysteresis.

Across the hypothetical sequence, the CCF lag is almost consistently decreasing, except for Obs-a, QPO frequency $F_+$ decreases from Obs-b to Obs-3, then increases, while the soft $F_{\rm QPO}$ {trough occurs} at Obs-1.
It seems that the soft $F_{\rm QPO}$ leads the long-term evolution, with the QPO frequency, QPO time lag (also $F_+$), and CCF lag responding to the call of QPO in succession.
This interconnected evolution and the corresponding models will be further explored in our forthcoming work {(Xia et al., in preparation)}.

\section{Conclusions}\label{sec5}

We perform wavelet analysis to explore the short-term variability of the QPO properties using high-quality observations with a cumulative exposure longer than 1000 ks.
Previous limited observational data indicates little correlation between the X-ray flux and the wavelet amplitude \citep{2010A&A...524A..26C}. However, in this study, we demonstrate the AM of the {hard} QPO and its strong correlation with the soft flux in a short timescale.
The conclusions are detailed as follows:

\begin{enumerate}
    \item The QPO signals are undergoing AM, both in the soft and hard bands. The timescales of this modulation are approximately 17 ks, corresponding exactly to the $\rm (5.25\pm 1.86) \times 10^{-5}\ Hz$ oscillation of the soft light curve, and close to the previously found FM timescale of 24 ks {within uncertainties}.

    \item The soft flux level is significantly higher when the hard QPO is present. They are highly correlated with an average CCF peak coefficient of {$0.61\pm 0.13$} and CCF lag of approximately 3 ks. This novel finding provides fresh insights into the potential connection between the components of the corona emitting soft and hard X-ray photons.

    \item The CCF lag between the soft flux and the hard QPO evolves across the observations, tracking closely and successively with the QPO time lag, the QPO frequency, and the fraction of time when the soft QPO is present. These QPO variation characteristics may share the same origin as the interconnected evolution between QPO frequency and time lag previously observed.

\end{enumerate}

\begin{acknowledgments}
This work is based on observations obtained with XMM-Newton, an ESA science mission with instruments and contributions directly funded by ESA Member States and NASA.
R.S.X., H.L., \& Y.Q.X. acknowledge support from the Strategic Priority Research Program of the Chinese Academy of Sciences (grant NO. XDB0550300), the NSFC grants (12025303 and 12393814), and the National Key R\&D Program of China (2023YFA1608100 and 2022YFF0503401).
\end{acknowledgments}

\bibliography{sample631}{}
\bibliographystyle{aasjournal}

\newpage
\appendix
\section{Averaged PSD}\label{secA}
In the previous work \citep{2024ApJ...961L..32X}, a weak but broad bulge in the PSD of the soft light curve is occasionally shown around $\rm (4\text{--}7) \times 10^{-5}\ Hz$ (e.g.{,} the Figure~2 in \citealt{2024ApJ...961L..32X}). However, the significance is quite low and no statistical tool can {assess it properly}.
The correlation of the QPO amplitude and the flux motivates us to confirm this weak signal. We average the PSDs of the 12 observations and fit the averaged PSD with a single power-law for simplicity (see Figure~\ref{figA1}). We see an evident bulge around the range of $\rm (5.25\pm 1.86) \times 10^{-5}\ Hz$, which is derived by simply fitting the residuals with a single Gaussian profile. We see the characteristic frequency of the QPO AM and the lower frequency of the soft flux are almost identical within one standard deviation.

\section{Correlation between Flux and QPO AM/FM}\label{secB}

We calculate the normalized CCF of the count rates with respect to both the QPO wavelet amplitude (represented by the integrated wavelet map in the QPO frequency range) and the QPO frequency (represented by the expectation of the QPO frequency concerning the wavelet amplitude in the QPO frequency range) for each case (see {Figures}~\ref{figB1}~and\ref{figB2}). The CCFs exhibit similar shapes across all cases in Figure~\ref{figB1}. Consistent with the findings in Figure~\ref{fig2}, the maximum peak correlation of {$0.61 \pm 0.13$} occurs between the soft count rate and the wavelet amplitude of the hard QPO (lower left panel in Figure~\ref{figB1}). 

However, this obvious correlation is not observed between the flux and the QPO frequency {(see Figure~\ref{figB2})}. The peak coefficients are much lower, and the CCFs show larger dispersion. No {apparent} long-term evolution of the CCF lags is observed, either.

\section{QPO Time Lag vs. Flux}\label{secC}

Similar to Figure~\ref{fig3}, we divide the light curves into two parts: one with positive lags between the soft and hard bands, and the other with negative lags. We then construct histograms of the soft and hard count rates, revealing a slight deviation between the two cases.
Given the significant evolution of the fraction of time with a positive time lag over the hypothetical sequence, we are inclined to believe it is influenced by long-term evolution, which will be detailed in our forthcoming work (Xia et al., in preparation).

\renewcommand{\thefigure}{A\arabic{figure}}
\setcounter{figure}{0}

\begin{figure*}
    \centering
    \includegraphics[scale=0.45]{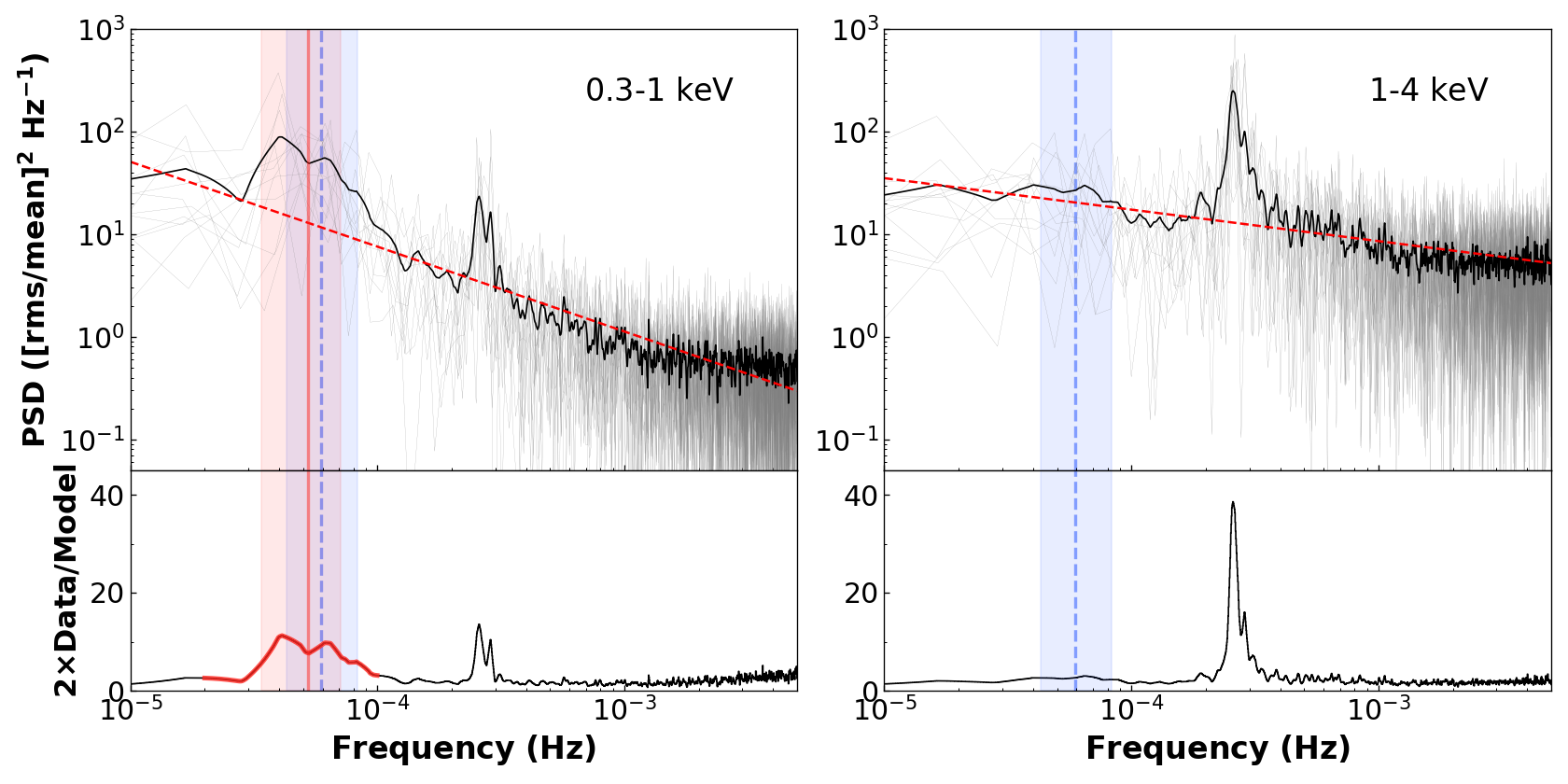}
    \caption{Average PSDs (black solid curves)  for RE J1034 + 396 in the soft band (0.3–1 keV) and the hard band (1–4 keV), respectively.  In the upper panels, the PSD for each observation is plotted in grey, and the red dashed lines depict the components of a power law. The bottom panels illustrate the ratios of the observed PSDs to the power law. The blue dashed lines and the shaded regions represent the timescale of $16.9 \pm 7.3$ ks of QPO AM in frequency, while the red solid line and the shaded regions represent the mean value and the standard deviation of the broad bulge of $(5.25\pm 1.86)\times 10^{-5}$ Hz, marked by the red thick solid curve of the soft band in the bottom panel.}
    \label{figA1}
\end{figure*}
\renewcommand{\thefigure}{B\arabic{figure}}
\setcounter{figure}{0}
\begin{figure*}
    \centering
    \includegraphics[scale=0.8]{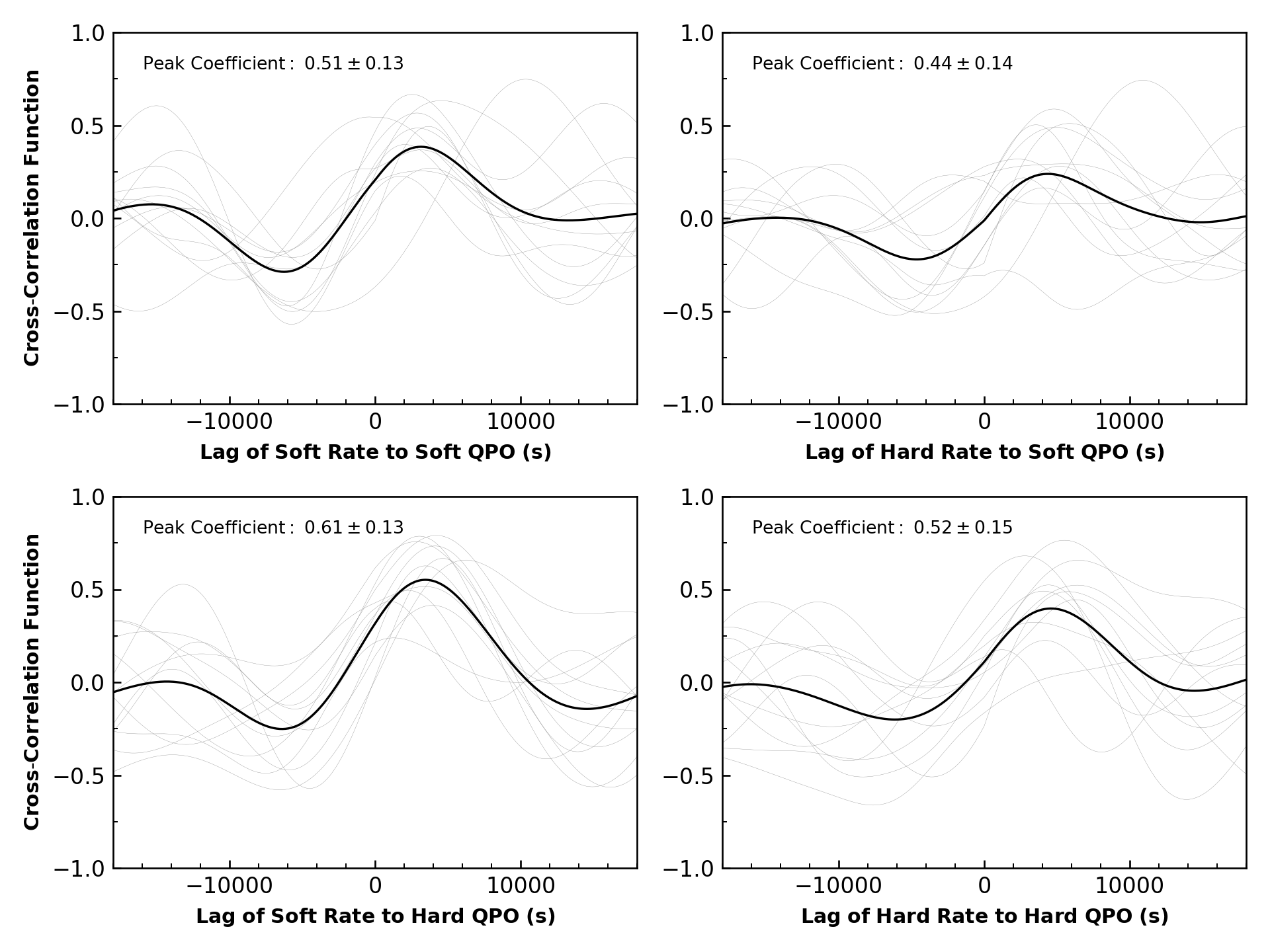}
    \caption{Normalized CCFs (grey lines) of the smoothed count rate for the soft (left panels) and hard (right panels) bands, and the QPO wavelet amplitude (represented by the integrated wavelet map in the QPO frequency range) for the soft (upper panels) and hard (lower panels) bands for each observation. The normalized CCFs for each observation are plotted as thin grey lines. The black solid lines represent the mean values of these grey lines, indicating the average CCF for all observations. The peak value {of the black solid line} is noted at the upper left of each panel.}
    \label{figB1}
\end{figure*}

\begin{figure*}
    \centering
    \includegraphics[scale=0.8]{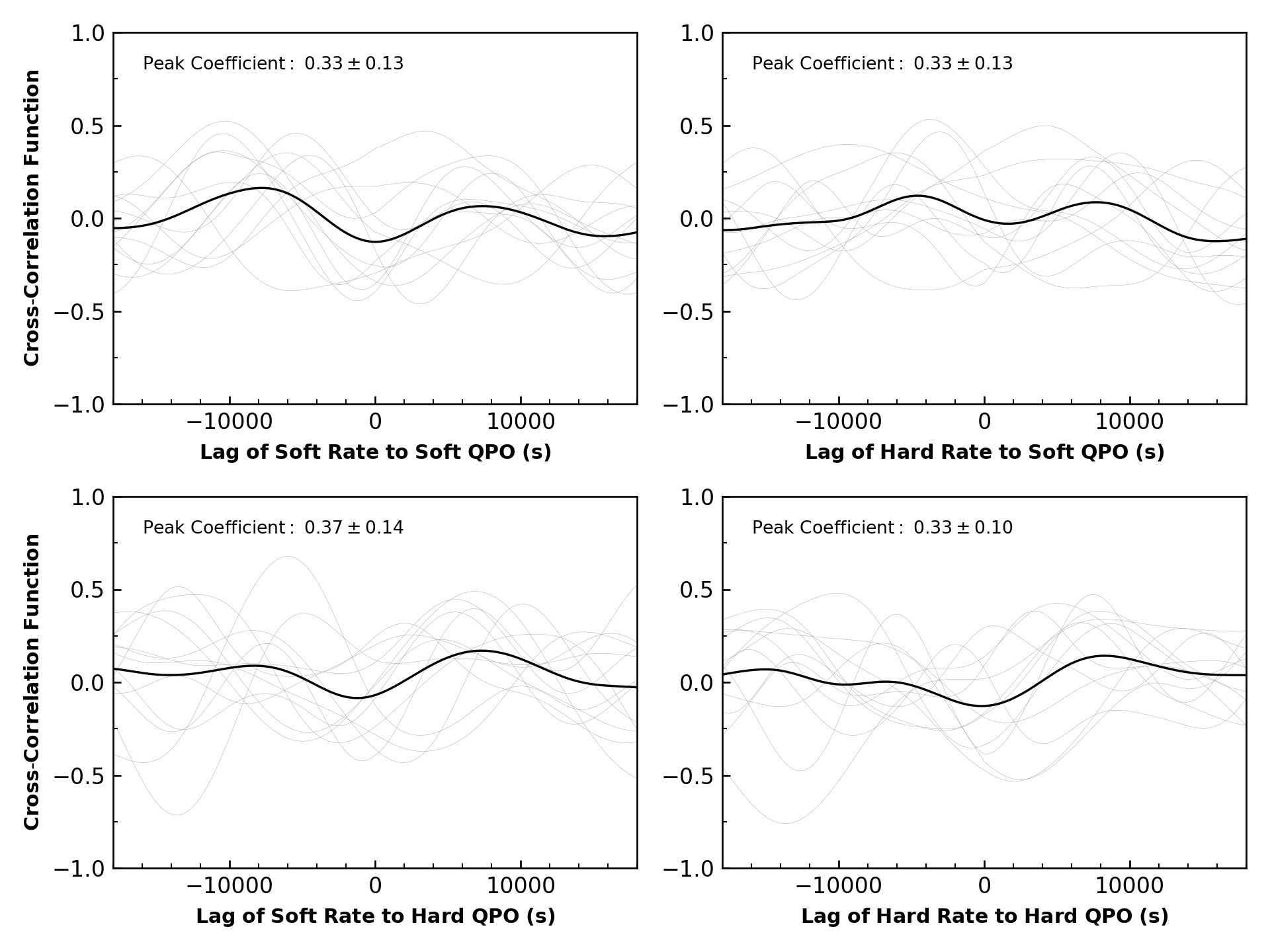}
    \caption{Same as Figure~\ref{figB1}, but for the flux (represented by the smoothed count rate) for the soft (left panels) and hard (right panels) bands, and the QPO frequency (represented by the expectation of the QPO frequency concerning the wavelet amplitude in the QPO frequency range) for the soft (upper panels) and hard (lower panels) bands.}
    \label{figB2}
\end{figure*}
\renewcommand{\thefigure}{C\arabic{figure}}
\setcounter{figure}{0}
\begin{figure*}
    \centering
    \includegraphics[scale=0.8]{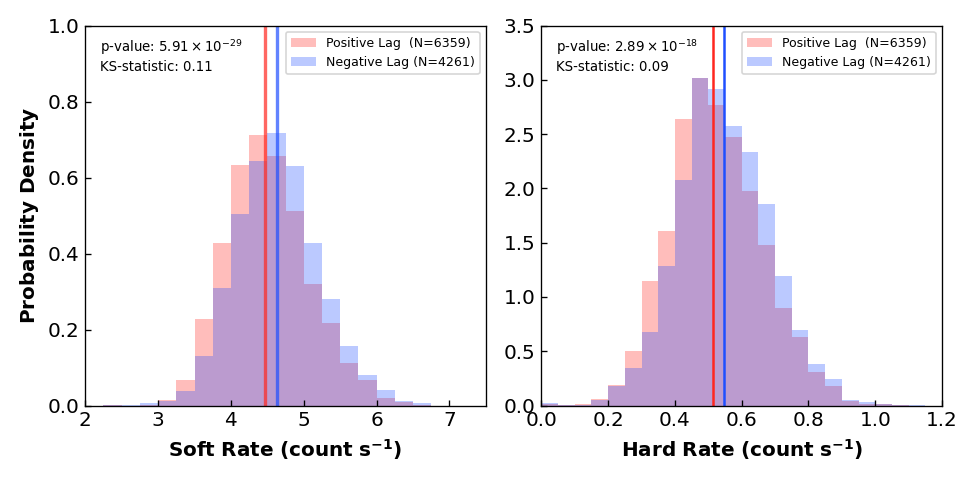}
    \caption{The distributions for the count rates of the soft (left {panel}) and hard bands (right {panel}) when the QPO time lag is positive (red) or negative (blue). The p-value and the KS-statistic are presented in the upper left corner of each panel.}
    \label{figC1}
\end{figure*}

\end{document}